\documentstyle[12pt]{article}

\def\thebibliography#1{\section*{
References}\list
  {\arabic{enumi}.}{\settowidth\labelwidth{#1}\leftmargin\labelwidth
    \advance\leftmargin\labelsep
    \usecounter{enumi}}
    \def\newblock{\hskip .11em plus .33em minus .07em}
    \sloppy\clubpenalty4000\widowpenalty4000
    \sfcode`\.=1000\relax}

\def\op#1{\mathop{\fam0 #1}\limits}

\newcommand{\beq}{\begin{equation}}
\newcommand{\eeq}{\end{equation}}
\newcommand{\ben}{\begin{eqnarray}}
\newcommand{\een}{\end{eqnarray}}
\newcommand{\be}{\begin{eqnarray*}}
\newcommand{\ee}{\end{eqnarray*}}
\newcommand{\bea}{\begin{eqalph}}
\newcommand{\eea}{\end{eqalph}}

\newcommand{\cL}{{\cal L}}

\newcommand{\bR}{{\bf R}}

\newcommand{\vt}{\vartheta}

\newcommand{\la}{\lambda}

\newcommand{\om}{\omega}
\newcommand{\Om}{\Omega}
\newcommand{\m}{\mu}

\newcommand{\w}{\wedge}

\newcommand{\dr}{\partial}
\newcommand{\ot}{\otimes}

\newenvironment{eqalph}{\stepcounter{equation}
\setcounter{equationa}{\value{equation}}
\setcounter{equation}{0}

\begin{eqnarray}}{\end{eqnarray}
\setcounter{equation}{\value{equationa}}}

\begin{document}
\hbox{}

\begin{center}
{\large \bf DEFORMATION QUANTIZATION IN COVARIANT  
\medskip 

HAMILTONIAN FIELD THEORY}
\bigskip

{\sc G.Sardanashvily}

Department of Theoretical Physics, Moscow State University

117234 Moscow, Russia

E-mail: sard@grav.phys.msu.su 
\end{center}

\bigskip

\begin{abstract}
The deformation star product of smooth functions on the momentum
phase space of covariant (polysymplectic) Hamiltonian field theory is
introduced. 
\end{abstract}
\bigskip

It is well known that, applied to field theory, the familiar
symplectic technique of mechanics takes the form of instantaneous Hamiltonian
formalism on an infinite-dimensional phase space. 
The finite-dimensional covariant Hamiltonian approach to field theory is
vigorously developed from the seventies in its multisymplectic and
polysymplectic variants, related to the two different Legendre
morphisms in the first 
order calculus of variations on fibre bundles (see [1-5] for a
survey).

Recall that, given a fibre bundle
$Y\to X$ coordinated by $(x^\la,y^i)$, a first order Lagrangian $L$ is
defined as a semibasic  density
\be
L=\cL\om: J^1Y\to\op\w^nT^*X, \quad \om=dx^1\w\cdots dx^n, \quad n=\dim X,
\ee
on the jet manifold $J^1Y$ of $Y$. 
Every Lagrangian $L$ yields the Legendre map of $J^1Y$ to the Legendre bundle 
\be
\Pi=\op\w^nT^*X\op\ot_YV^*Y\op\ot_YTX,
\ee
coordinated by $(x^\la,y^i,p^\la_i)$. It is provided 
with the canonical polysymplectic form
\be
\Om_Y =dp_i^\la\w dy^i\w \om\ot\dr_\la,
\ee
and is regarded as the polysymplectic momentum phase space of fields.

The multisymplectic momentum phase space of fields is 
the homogeneous Legendre bundle
\be
Z_Y= T^*Y\w(\op\w^{n-1}T^*X),
\ee
coordinated by $(x^\la,y^i,p^\la_i,p)$ and equipped with the 
canonical multisymplectic form
\be
d\Xi_Y= dp\w\om + dp^\la_i\w dy^i\w\om_\la. 
\ee
The
relationship  between  polysymplectic and multisymplectic phase spaces is
given by the exact sequence
\be
0\to\Pi\op\times_X\op\w^nT^*X\hookrightarrow Z_Y\to\Pi\to 0, 
\ee
where $Z_Y\to \Pi$ is a one-dimensional affine bundle. Given a
section
$h$ of $Z_Y\to\Pi$, the pull-back $h^*\Xi_Y$ is a
polysymplectic Hamiltonian form on $\Pi$. 

It seems natural to generalize a Poisson
bracket in symplectic mechanics to polysymplectic or multisymplectic
manifolds in order to obtain the covariant canonical
quantization of field theory. Different variants of such a bracket have
been suggested (see the recent works [6,7]). The main
difficulty is that the bracket must be globally defined, i.e.,
maintained under the bundle coordinate transformations.

Let us assume that a fibre bundle $Y\to X$ is affine. All realistic field
models are of these type, and only vector and affine fields are quantized.
One can introduce the bilinear differential operator 
\be
\Delta_\m: (f,f') \mapsto f{\op\Delta^\leftrightarrow}_\m f'=\dr^i_\m
f\dr_i f'- \dr_i f\dr^i_\m f' 
\ee
on smooth functions $f,g\in C^\infty(\Pi)$ on $\Pi$ [8]. Let $X$ be
provided with a nowhere vanishing vector field $\vt$. This condition implies
the existence of a pseudo-Riemannian metric $g$ on $X$ such
that integral curves of $\vt$ are time lines with respect to $g$. Let us
consider the operator 
\ben
&& \Delta: C^\infty(\Pi)\times C^\infty(\Pi)\to C^\infty(\Pi),\nonumber\\
&& \Delta: (f,f')\mapsto f{\op\Delta^\leftrightarrow}f'=
f[{\op\Delta^\leftrightarrow}_\m \vt^\m]f'. \label{pr3}
\een
If $X=\bR$, then $\vt=\dr_t$ and $\Delta$ (\ref{pr3}) is the canonical Poisson
bracket in Hamiltonian time-dependent mechanics [9,10].

Therefore, by analogy with the well-known Moyal product on symplectic
manifolds, 
one can introduce
the associative star product 
\be
f*f'=f\exp\{-\frac{i\hbar}{2} {\op\Delta^\leftrightarrow}\} f'
\ee
and the bracket 
\be
[f,f']=f*f'-f'*f
\ee
of smooth functions on $\Pi$. This bracket provides deformation
quantization of the momentum phase space of classical fields $\Pi$.
Similarly to the familiar quantization in instantaneous Hamiltonian
field theory,  
this deformation quantization is tangential to the integral curves
of the vector field $\vt$, i.e., to the time lines.

\end{document}